# Extreme Nonreciprocity with Spatio-Temporally Modulated Metasurfaces

Andrew E. Cardin[1, 2], Sinhara R. Silva[1], Shai. R. Vardeny[1], Willie J. Padilla[2], Avadh Saxena[3], Antoinette J. Taylor[1], Wilton J. M. Kort-Kamp[3], Hou-Tong Chen[1], Diego A. R. Dalvit[3]*, and Abul K. Azad[1]*

[1] *Center for Integrated Nanotechnologies, Los Alamos National Laboratory, Los Alamos, New Mexico 87545, USA*

[2] *Department of Electrical and Computer Engineering, Duke University, Durham, North Carolina 27708, USA*

[3] *Theoretical Division, Los Alamos National Laboratory, Los Alamos, New Mexico 87545, USA*

* Corresponding authors: A.A.**:** aazad@lanl.gov and D.D.: dalvit@lanl.gov

**Emerging photonic functionalities are mostly governed by the fundamental principle of Lorentz reciprocity. Lifting the constraints imposed by this principle could circumvent deleterious effects that limit the performance of photonic systems. A variety of approaches have recently been explored to break reciprocity, yet most efforts have been limited to confined photonic systems. Here, we propose and experimentally demonstrate a spatio-temporally modulated metasurface capable of extreme breakdown of Lorentz reciprocity. Through tailoring the momentum and frequency harmonic contents of the scattered waves, we achieve dynamical beam steering, reconfigurable focusing, and giant free-space optical isolation exemplifying the flexibility of our platform. We develop a generalized Bloch-Floquet theory which offers physical insights into the demonstrated extreme nonreciprocity, and its predictions are in excellent agreement with experiments. Our work opens exciting opportunities in applications where free-space nonreciprocal wave propagation is desired, including wireless communications and radiative energy transfer.**

The electromagnetic Lorentz reciprocity theorem[1-4] states that in a linear, time-independent system with symmetric constitutive optical tensors, the ratio between received and transmitted fields are the same for forward and time-reversed propagation directions. While most electromagnetic and photonic devices operate under this principle, there are circumstances in which reciprocity has undesirable effects, *e.g.*, photovoltaic cells re-emitting absorbed solar energy or antennas listening to their own echo. In order to break reciprocity, any of the conditions assumed by the Lorentz theorem must be violated. The use of magneto-optical media breaks time-inversion symmetry and results in an asymmetric scattering matrix, thereby accomplishing optical isolation[5,6]; however, bulky magnets are required for external bias renders this conventional approach infeasible for systems integration. Incorporating nonlinear materials and using the electric field self-biasing effect can also result in optical isolation[7-9], but the degree of nonreciprocity is power-dependent and typically requires significant interaction lengths. Breakdown of Lorentz reciprocity has also been demonstrated using spatio-temporally modulated waveguides[10-12] and leaky-wave antennas[13-14]. The space-time modulation approach can be particularly attractive when applied in metasurface platforms due to its advantages in reduced size, improved integrability, and for attaining extreme nonreciprocal behavior as we demonstrate in this work.

The advent of metasurfaces[15-20] has allowed enhanced light-matter interactions within ultra-thin structures, enabling tailored scattering amplitude, phase, and polarization, as well as facilitating the integration of functional materials to accomplish active control[21-25]. Spatially varying phase profiles in judiciously designed static gradient metasurfaces have provided a powerful means to manipulate the momentum harmonic contents of scattered light, ushering in a novel class of flat optics. In addition, active metasurfaces have been shown to be capable of either switching wave-front profiles whenever needed[26-28], or modulating in time the amplitude and phase of scattered light[29,30]. However, a fully tailored electromagnetic response requires metasurfaces that can continuously alter their scattering properties simultaneously in *time and space*. These functionalities can be achieved in spatio-temporally modulated metasurfaces[31] (STMMs), which have the potential to revolutionize fundamental and applied photonics, including nonreciprocity[32,33], through on-demand control of frequency and momentum harmonic contents of scattered light. For example, an STMM can focus a collimated beam (Fig. 1a), but the time-reversed process is not allowed (Fig. 1b). At microwave frequencies, STMMs can be realized with sub-wavelength metasurface resonators embedded with active elements, such as positive-intrinsic-negative diodes or varactors, which are locally modulated through programmable space-time voltage biases. This concept has been employed for frequency-multiplexed directional scattering within waveguide systems containing time-varying Huygens' metadevices[34]. For free-space waves, spatio-temporally modulated coding metasurfaces have been recently exploited for active and nonreciprocal beam steering[35,36]; however, only weak nonreciprocity limited to the frequency-domain was accomplished. A truly multifunctional STMM that can achieve nonreciprocal propagation of *arbitrary wave-fronts*, both in the frequency and spatial domains, and that can even reach *extreme breakdown* of Lorentz reciprocity, has not been experimentally demonstrated to date.

Here, we present a spatio-temporally modulated metasurface reflectarray capable of dynamically imprinting arbitrary wave-fronts of free-space electromagnetic waves, and demonstrate maximum breakdown of Lorentz reciprocity by reflecting an incident beam into far-field radiation in forward scattering but into near-field surface waves in reverse scattering. We exemplify the flexibility of our STMM platform by achieving dynamical beam steering, reconfigurable focusing, and giant free-space optical isolation due to nonreciprocal excitation of surface waves. To model the experimental results, an analytical generalized Bloch-Floquet theory valid for arbitrary phase distributions is developed, which successfully predicts various phenomena measured in the extreme nonreciprocity regime. This work will strongly impact emerging technologies benefiting from free-space dynamical wave-front shaping and nonreciprocity, including adaptive optics, Doppler-like frequency translation, echo-immune antennas, and isolated on-chip communications enabled by robust one-way coupling to surface waves.

**Results**

**Experimental design.** A conceptual illustration of an STMM is shown in Fig. 1a,b, where each unit cell contains active elements that can be independently modulated to impart an arbitrary two-dimensional (2D) phase profile in reflection. In this work, we consider STMMs working at microwave frequencies and varactors as the active elements. Each varactor at position $r$ on the STMM is subjected to a time-dependent voltage modulation $V(r,t) = V_{\text{op}} + \Delta_V \sin[\varphi(r) - \Omega t]$, where $V_{\text{op}}$ is the operating voltage (or off-set bias), $\Delta_V$ is the voltage modulation amplitude, $\varphi(r)$ is the spatial phase distribution, and $\Omega$ is the modulation frequency. The voltage modulation translates into a corresponding modulation of the capacitance $C_{\text{var}}(r,t) = C_{\text{op}} + \Delta_C \sin[\varphi(r) - \Omega t]$ as long as a linear relationship between $C_{\text{var}}$ and $V$ holds near $V_{\text{op}}$.

We design and fabricate an STMM on a commercially available double copper clad FR-4 substrate (Fig. 1c,d). The resonators incorporate two varactor diodes and two capacitors in order to keep a small form factor of the unit cell while providing the required tunability. For simplicity, we restrict the modulation to one dimension by connecting resonators in each column, resulting in spatial modulation among different columns, *i.e.*, along the *x*-direction, but invariant among rows, *i.e.*, *y*-direction (phase delays along this direction can be ignored for $\Omega \ll 2\pi c/L_y$, where $c$ is the speed of light in vacuum and $L_y$ is the lateral size of the STMM along *y*). Each column is modulated with a desired voltage and phase using a programmable multi-output function generator. Relative phase shifts among channels can be induced by simultaneous triggering of the system, thereby generating arbitrary digitally constructed waveforms (see Methods and Supplementary Fig. 1).

The reflectivity of the unmodulated metasurface is characterized inside an anechoic chamber at a 10° incidence angle as a function of frequency for various static bias voltages uniformly applied to all varactors. Figure 2a shows measurements for *p*-polarized waves, with the magnetic field component along the *y*-direction. At the operating voltage $V_{\text{op}} = 2$ V a resonance appears around 6.6 GHz. The optimal working frequency for the modulated STMM, however, is not exclusively determined by the intrinsic resonances of the unmodulated metasurface, but results from an interplay between the unit cell design and the amplitude and phase of the modulation. Indeed, working away from the 6.6 GHz resonance increases the modulation efficiency (see Supplementary Fig. 4c) and thereby the conversion into frequency harmonics (see Supplementary Fig. 5); however, moving too far from the resonance degrades phase dispersion and breaks the linear $C_{\text{var}}$ vs $V$ relationship (see Supplementary Fig. 4d). For our designed STMM and modulation protocol ($\Delta_V = 1$ V, $\Omega = 2\pi \times 50$ kHz), we choose our operational frequency as $\omega_{\text{in}} = 2\pi \times 6.9$ GHz ($\lambda_{\text{in}} = 4.3$ cm), which is a good compromise between reflection efficiency, phase dispersion, and linearity.

To test the nonreciprocal functionalities of our spatio-temporally modulated metasurface, we first perform forward scattering measurements by dynamically imprinting appropriate phase delays $\varphi(\boldsymbol{r})$ among the STMM columns. When applying a linear phase delay, we measure the steered beam as a function of scanning angle and frequency in the far-field, while for focusing we apply a parabolic phase profile for a selected frequency harmonic and map the resulting spatial distribution of the scattered power. Subsequently, the corresponding reverse scattering measurements are performed by effectively time-reversing the output wave-fronts of the forward processes, and measuring the far-field scattered power.

**Theoretical formulation.** We describe the response of the full array of sub-wavelength resonators with an effective 2D conductivity $\sigma(r, \omega; t) = \sigma_{\mathrm{op}}(\omega) + \Delta\sigma(\omega) \sin[\varphi(r) - \Omega t]$. Here, $\sigma_{\mathrm{op}}(\omega)$ is the unmodulated complex conductivity at the operating voltage $V_{\mathrm{op}}$ and $\Delta\sigma(\omega)$ is the complex modulation amplitude. We extract $\sigma_{\mathrm{op}}(\omega)$ and estimate $\Delta\sigma(\omega)$ from reflectivity measurements and standard Fresnel equations for multilayered systems (see Supplementary Fig. 4). When an incident plane wave $\boldsymbol{E}_\xi^{\mathrm{in}}(\boldsymbol{r}, z, t) = \hat{\boldsymbol{e}}_\xi^-(\boldsymbol{k}_{\mathrm{in}}) E^{\mathrm{in}} \times e^{i(\boldsymbol{r}\cdot\boldsymbol{k}_{\mathrm{in}} - z\, k_{\mathrm{zv,in}} - \omega_{\mathrm{in}} t)}$ with momentum $(\boldsymbol{k}_{\mathrm{in}}, -k_{\mathrm{zv,in}}\,\hat{\boldsymbol{z}})$ impinges on the STMM, the reflected field for an arbitrary phase distribution $\varphi(\boldsymbol{r})$ on the STMM is given as

$$\boldsymbol{E}_\xi^{\mathrm{refl}}(\boldsymbol{r}, z, t) = E^{\mathrm{in}} \sum_{n=-\infty}^{\infty} \int \frac{d^2\boldsymbol{k}}{(2\pi)^2}\, \hat{\boldsymbol{e}}_\xi^+(\boldsymbol{k}_n)\, \widetilde{\mathcal{E}}_{\xi,n}[\boldsymbol{k}_{\mathrm{in}}, \boldsymbol{k}_n, \omega_n; \boldsymbol{r}]\, e^{i(\boldsymbol{r}\cdot\boldsymbol{k}_n + z\, k_{\mathrm{zv},n} - \omega_n t)}. \qquad (1)$$

Here, $\hat{\boldsymbol{e}}_\xi^\pm(\boldsymbol{k})$ are the $\xi = s, p$ polarization unit vectors with '+' and '−' corresponding to waves propagating in the positive and negative z directions, respectively; $\omega_n = \omega_{\mathrm{in}} + n\,\Omega$ and $\boldsymbol{k}_n = \boldsymbol{k} + n\,\nabla\varphi(\boldsymbol{r})$ are frequency and momentum harmonics, and $k_{\mathrm{zv},n} = \sqrt{(\omega_n/c)^2 - \boldsymbol{k}_n^2}$. To compute the reflected field, we develop an analytic Bloch-Floquet approach based on a local derivative expansion of $\varphi(\boldsymbol{r})$, valid for smooth but otherwise general phase distributions. The field amplitude $\widetilde{\mathcal{E}}_{\xi,n}$ is a functional of both $\varphi(\boldsymbol{r})$ and its gradient $\nabla\varphi(\boldsymbol{r})$,

$$\widetilde{\mathcal{E}}_{\xi,n} = R_{\xi,n}(\boldsymbol{k}, \omega_{\mathrm{in}}; \boldsymbol{r}) \left\{ \delta_{n,0}\, \delta(\boldsymbol{k} - \boldsymbol{k}_{\mathrm{in}}) + \frac{(1 - \delta_{n,0})\, \mathrm{FT}[e^{i\,\mathrm{sg}(n)\varphi(\boldsymbol{r})}]}{e^{i\,\mathrm{sg}(n)[\varphi(\boldsymbol{r}) - \boldsymbol{r}\cdot\nabla\varphi(\boldsymbol{r})]}} \Lambda_{\xi,n}[\boldsymbol{k}_{\mathrm{in}}, \boldsymbol{k}_n, \omega_n; \boldsymbol{r}] \right\}, (2)$$

Here, $R_{\xi,n}(\mathbf{k}, \omega_{\text{in}}; \mathbf{r})$ is the reflection coefficient corresponding to the scattering process $(\mathbf{k}, \omega_{\text{in}}) \to (\mathbf{k}_n, \omega_n)$ at position $\mathbf{r}$, the 2D Fourier transform $\text{FT}[e^{i\,\text{sg}(n)\varphi(\mathbf{r})}]$ is evaluated at $\mathbf{k} - \mathbf{k}_{\text{in}} + \text{sg}(n)\nabla\varphi(\mathbf{r})$, and $\Lambda_{\xi,n}$ is defined in the Methods section. Finally, we mention that when the modulation takes place only over a finite-sized region of the metasurface, the reflected field is still given by Eq. (1) after the Fourier transform in Eq. (2) is replaced by $\text{FT}[\tilde{\theta}(\mathbf{r})\, e^{i\,\text{sg}(n)\varphi(\mathbf{r})}]$, where $\tilde{\theta}(\mathbf{r}) = 1$ if $\mathbf{r}$ belongs to the modulated region and is zero otherwise (see Supplementary Information).

**Dynamical wave-front shaping.** In the simplest case of a linear phase distribution $\varphi^{\text{steer}}(\mathbf{r}) = \boldsymbol{\beta} \cdot \mathbf{r}$ required for beam steering, the amplitudes are $\tilde{\mathcal{E}}_{\xi,n}^{\text{steer}} = (2\pi)^2 R_{\xi,n}^{\text{steer}}(\mathbf{k}, \omega_{\text{in}}; \mathbf{0})\, \delta(\mathbf{k} - \mathbf{k}_{\text{in}})$. In this case the derivative expansion is exact, and the reflected field is given by a single sum over discrete frequency and momentum harmonics for the scattering processes $(\mathbf{k}_{\text{in}}, \omega_{\text{in}}) \to (\mathbf{k}_{\text{in}} + n\boldsymbol{\beta}, \omega_{\text{in}} + n\Omega)$. The azimuthal and polar steering angles are respectively given by $\cos\phi_n = \mathbf{k}_{\text{in}} \cdot (\mathbf{k}_{\text{in}} + n\boldsymbol{\beta})/[|\mathbf{k}_{\text{in}}||\mathbf{k}_{\text{in}} + n\boldsymbol{\beta}|]$ and $\sin\theta_n = |\mathbf{k}_{\text{in}} + n\boldsymbol{\beta}|c/\omega_n$. For large enough momentum "kicks" $n\boldsymbol{\beta}$, $|\sin\theta_n|$ can become larger than 1 ($k_{zv,n}$ purely imaginary) resulting in surface waves, which we later utilize to achieve giant nonreciprocity. We first demonstrate the multifunctional beam manipulation capability of our STMM by imprinting arbitrary beam steering phase distributions (Fig. 2b). We vary the phase gradient $\boldsymbol{\beta}$ along the *x*-direction and show beam steering of the +1 harmonic from +40° to -40°. We observe high quality beam steering, with low sidebands, a consistent amplitude for a wide range of angles, and ~0.1% conversion efficiency.

To test the flexibility of our STMM for imprinting arbitrary reflection phases, we demonstrate dynamical focusing, a functionality that is relevant, *e.g.*, in compact satellite technologies. The required phase distribution to focus a specific frequency harmonic $\bar{n} \neq 0$ at an arbitrary focal point $\boldsymbol{R}_f = (x_f, y_f, z_f)$ is

$$\varphi_{\bar{n}}^{\text{focus}}(\boldsymbol{k}_{\text{in}}, \omega_{\text{in}}; \boldsymbol{r}) = -\frac{1}{\bar{n}}\left\{\frac{\omega_{\bar{n}}}{c}\left(|\boldsymbol{r} - \boldsymbol{R}_f| - z_f\right) + \boldsymbol{k}_{\text{in}} \cdot (\boldsymbol{r} - \boldsymbol{R}_f)\right\}. \tag{3}$$

All other harmonics $n \neq \bar{n}$ do not result in perfect focusing, and the case $n=0$ always undergoes specular reflection in STMMs. In contrast to the steering case, the momentum kicks $n\nabla\varphi_{\bar{n}}^{\text{focus}}$ vary as a function of position and the amplitude $\widetilde{\mathcal{E}}_{\xi,n}^{\text{focus}}$ is not proportional to a delta function. These properties hold for any nonlinear phase distribution $\varphi_{\text{NL}}(\boldsymbol{r})$. Importantly, the phase of the reflected field in Eq. (1) arises from an intricate interplay among the phases of each plane-wave component in the momentum integral. Only for the $n = \bar{n}$ frequency harmonic does the resulting reflection phase turn out to be precisely given by the focusing distribution Eq. (3) (see Supplementary Information). Using our theoretical approach, we model an infinite-sized STMM subjected to our focusing modulation protocol, that produces a 1D focal line along the $y$-direction, and compute the field distribution for the $n = \bar{n} = +1$ harmonic (Fig. 2c).

In Fig. 2d and Fig. 2f we show the experimental results demonstrating the dynamical focusing capability of our STMM (see Methods). Unlike beam steering, which is largely aperture agnostic, focusing requires consideration of the metasurface's overall dimensions. With a surface in the electric field direction measuring $L_x = 19$ cm ($\approx 4.4\,\lambda_{\text{in}}$), a short focal length is necessary for measurable gain, otherwise weak focusing effects are indistinguishable from the usual plane wave like character of the harmonics. For the on-axis focusing case, it is apparent that focusing has been achieved by examination of the increased power and rapid falloff of the signal. At distances $\ell < 12$ cm (measured from the STMM centre along the focal axis) we observe interference due to near-field coupling between the scanning monopole and the STMM, evident in Fig. 2d. For the off-axis experiment, we measure a much tighter focus due to minimized interference and shadowing effects. To evaluate the quality of the focusing we estimate the gain along the focal axis. We define gain as the ratio between the power of the focused beam at frequency $\omega_{+1}$ and that of a beam steered at the same frequency $\omega_{+1}$ into the direction of the focal axis. Figure 2e compares the experimental gain for the off-axis case with the theoretical predictions using the extension of our theory approach to finite-sized STMMs. The experimental data and the theory display the same shape along the focal axis, showing strong agreement. For $x_f = 6$ cm and $z_f = 15$ cm, the expected focal length is $\ell_f = 16$ cm. However, due to the finite size of our metasurface, the gain peaks around $\ell = 13$ cm, both in experiment and theory (Fig. 2e). At this point, theory predicts a gain of 5.24 dB while the measured value is $(5.8 \pm 0.6)$ dB.

**Extreme breakdown of Lorentz reciprocity.** We turn our attention to nonreciprocal excitation of surface waves both in beam steering and focusing. In Fig. 3a we report the $n = +1$ beam steered to $\theta_{+1} \approx +18°$ ($\beta_x = 44$ m$^{-1}$) from a normally incident plane wave (see Fig. 3c for the theory results). In the reverse experiment we send the time-reversed $n = +1$ beam $(-\mathbf{k}_{\text{in}} - \boldsymbol{\beta}, \omega_{\text{in}} + \Omega)$ onto the STMM (see Methods). A frequency down-conversion and a momentum up-conversion must take place in order for the scattering process to be reciprocal. However, this is not permitted due to the travelling-wave nature of the modulation, which for frequency down-conversion results in an output $(-\mathbf{k}_{\text{in}} - 2\boldsymbol{\beta}, \omega_{\text{in}})$, clearly breaking reciprocity in the space-domain (direction of propagation). The measured reverse reflection is shown in Fig. 3b, with a reflection angle $\approx -36°$ (in contrast to the 0° reciprocal case) in excellent agreement with the theory prediction of -37.5° (Fig. 3d) (other possible reverse pathways breaking reciprocity in both space- and frequency-domains are discussed in the Supplementary Information). When the magnitude of the phase gradient surpasses a certain threshold ($|\boldsymbol{\beta}| > \omega_{\text{in}}/2c = 72.31$ m$^{-1}$ for $\mathbf{k}_{in} = 0$, corresponding to a forward reflection angle of $+30°$), the reverse scattered beam does not reflect off the STMM at all. Instead, surface waves are launched on the metasurface as $k_{zv,-1}^{REV,steer} = i[|\mathbf{k}_{in} + 2\boldsymbol{\beta}|^2 - (\omega_{in}/c)^2]^{1/2}$ is purely imaginary, resulting in perfect free-space optical isolation. In Fig. 3e,f we present nonreciprocity measurements for $\beta_x = 96$ m$^{-1}$ ($\theta_{+1} \approx +43°$), and in Fig. 3g,h the corresponding theory calculations. By scattering an incident beam into far-field collimated radiation in the forward process but into near-field evanescent modes in the reverse case, we demonstrate *extreme breakdown* of Lorentz reciprocity.

We now report giant optical isolation in dynamical focusing. In the forward process a plane wave ($\mathbf{k}_{in}, \omega_{in}$) can focus at frequency $\omega_{+1}$ using the $\varphi_{\bar{n}=+1}^{focus}(\mathbf{k}_{in}, \omega_{in}; \mathbf{r})$ phase profile. Figure 4a shows the measured power profile of the forward reflected off-axis focused beam as a function of $\ell$ and the angle $\alpha$ between the surface normal and the detector. We investigate the reflected field at $\omega_{in}$ after the focused beam at $\omega_{+1}$ is time-reversed in order to probe nonreciprocity in the spatial-domain. In order to grasp some physical insight, we show in Fig. 4b a snapshot of the electric field distribution for the particular scattering process ($-\mathbf{k}_{in} - \nabla\varphi_{\bar{n}=+1}^{focus}, \omega_{in} + \Omega) \rightarrow (-\mathbf{k}_{in} - 2\nabla\varphi_{\bar{n}=+1}^{focus}, \omega_{in}$) for the infinite-sized STMM, similarly to what is done above for nonreciprocal beam steering (see Methods). Also shown is the corresponding time-averaged Poynting vector. The reverse scattered field does not collimate back into the direction of the original input beam, but instead shows a "ribbon-like" region (centered at $x_f$ and of width $2z_f/\sqrt{3}$ for the chosen process) containing propagative modes ($k_{zv,-1}^{REV,focus} = [(\omega_{in}/c)^2 - |\mathbf{k}_{in} + 2\nabla\varphi_{\bar{n}=+1}^{focus}|^2]^{1/2}$ is real) that radiate away from the metasurface, and is flanked by a region of surface waves with purely imaginary $k_{zv,-1}^{REV,focus}$ (these latter modes move along the STMM and weakly manifest in Fig. 4b due to their strong decay). Other choices of scattering processes also present the same nonreciprocal characteristics. The total reverse scattered field results in maximum breakdown of Lorentz reciprocity.

For a finite-sized STMM our calculations indicate that a similar physics takes place, the most important difference being that the propagative waves are asymmetrically radiated from the edges of the STMM at angles $\alpha_{\pm}^{\text{th}}$. In Fig. 4c,d we experimentally validate these theoretical predictions by exciting the metasurface with a monopole antenna located at the focal point and emitting at frequency $\omega_{+1}$, and detect the far-field scattered radiation from the STMM at frequency $\omega_{\text{in}}$. The measured power map bears little resemblance to the original input plane wave. It shows a concave "wedge-shaped" region with low radiated power at the center, two peaks at asymmetric angular locations, and an outer region of negligible radiated power (displaying similarities to the Poynting vector depicted in Fig. 4b). All these properties of the reverse scattered field are in stark contrast to the input Gaussian profile for the forward process and in excellent agreement with the theory calculations for finite-sized STMMs. In particular, the differences in the measured angular positions of the peaks ($\alpha_{+}^{\text{th}} \neq -\alpha_{-}^{\text{th}}$) and in their strengths are due to the inherent asymmetry of off-axis focusing in finite-sized STMMs (see Supplementary Information). The measurements reported in Fig. 4c,d constitute the first experimental demonstration of extreme nonreciprocal focusing functionality using STMMs.

In contrast to the case of STMMs modulated with a linear phase $\varphi_{\text{L}}(\boldsymbol{r}) = \boldsymbol{\beta} \cdot \boldsymbol{r}$, for which surface waves in reverse scattering are launched only when its gradient surpasses a certain threshold, no such constraint exists on any nonlinear phase gradient $\nabla \varphi_{\text{NL}}(\boldsymbol{r})$ dynamically imprinted on the metasurface. In this sense, extreme breakdown of Lorentz reciprocity is a robust property occurring in space-time metasurfaces under arbitrary nonlinear spatial modulations.

**Discussion**

In summary, we have introduced an STMM platform for extreme breakdown of Lorentz reciprocity. This approach enables dynamical arbitrary wave-front shaping by offering substantial flexibility in the manipulation of frequency-momentum harmonic contents of free-space electromagnetic waves when compared to standard phase-gradient and nonlinear optical systems. We have also developed an analytical approach to model STMMs with arbitrary phase distributions and its predictions are in excellent agreement with experiment, shedding light on the underlying physics of STMMs with complex space-time modulations without resorting to full-wave numerical simulations. Maximum breakdown of Lorentz reciprocity can occur for any kind of linear or nonlinear spatial phase distributions dynamically imprinted on the metasurface, highlighting that extreme nonreciprocity is a generic property of STMMs. Furthermore, by independently addressing each individual resonator, our platform can be extended to achieve fully three-dimensional dynamical wave-front shaping. The simple sinusoidal modulation chosen in this work can be generalized to more complex modulation protocols, opening opportunities for efficiency improvement, momentum-frequency harmonic mode selectivity, and advanced wave-front manipulation. The fully customizable response of our metasurface is a new architecture for compact and flat multifunctional optical components with built-in isolators, reducing stringent size, weight, and power requirements for wireless communication and remote sensing. Additionally, such a spatio-temporally modulated platform can be used to compensate for Doppler shifts induced by relative motion in inter-satellite and Earth-to-satellite communications. Extension of STMMs to the THz and IR frequency range is possible by use of alternative materials and active elements, including back-gate modulated graphene-based resonators, and electro-optic and photo-acoustic media. Finally, our demonstration of maximum violation of Lorentz reciprocity for arbitrary wave-fronts supports emerging technologies benefiting from free-space optical isolation, such as nonreciprocal wireless information transmission.

**Acknowledgments**

We are grateful to the ISR Fab-Lab at LANL for mounting the varactors on the metasurface. D.D. thanks Lukasz Cincio for computational assistance. Research presented in this paper was supported by the Laboratory Directed Research and Development program of Los Alamos National Laboratory under project number 20180062DR. Part of this work was performed at the Center for Integrated Nanotechnologies, a U.S. Department of Energy, Office of Basic Energy Sciences user facility.  LANL is operated by Triad National Security, LLC, for the National Nuclear Security Administration of the U.S. Department of Energy (Contract No. 89233218CNA000001).

**Author contributions**

A.A., H.T.C., D.D., and W.K.K. conceived the idea of this work. A.A. and A.C. designed the experiments. A.C. and S.S. fabricated samples and performed the measurements. S.V. designed the electronic control system. D.D. and W.K.K. developed the theory and conducted numerical simulations. A.A., A.C., H.T.C., D.D., W.K.K., W.P., A.S., and A.T. contributed to the writing of the paper. A.A. and D.D. supervised the entire study.

**Competing interests**

The authors declare no competing interests.

**Additional information**

**Supplementary Information** is available for this paper.

**Correspondence and requests for materials** should be addressed to A.A. or D. D.

**Methods**

**Device fabrication.** The metasurface of dimensions $L_x = L_y = 19$ cm was fabricated using 12"x12" double sided FR-4 circuit boards, with a thickness of 1.6 mm and 1oz copper. The top side of the board was milled yielding an array of unit cells on one side, and a continuous ground plane separated by the FR-4 spacer. After milling, the varactors (SMV1405) and capacitors (SMD 0.6pF) were added by reflow soldering and electrical connections were made to the columns for external control. The voltage signals being fed into each column of the STMM were sourced from a high-density signal generator that is installed on a National Instruments PCI eXtensions for Instrumentation (PXIe) chassis. Each waveform was digitally constructed and uploaded to the onboard buffer of the control system. Both the PXIe chassis and the signal generators have synchronized sample clocks, thus preventing relative phase drift in time and enabling the triggering of all channels to within 10 ns. See Supplementary Fig. 1a for a photograph of the fabricated sample with control lines.

**Measurements.** The fabricated metasurface was characterized inside an anechoic chamber using a broad-band horn antenna (SAS-571) and a quarter wave monopole antenna (see Supplementary Figs. 1b,d for a schematic of the experimental set-ups for dynamical steering and focusing, and Supplementary Figs. 1c,e for nonreciprocal beam steering and focusing). A vector network analyzer (VNA - Agilent N5230A) was used for data collection. First, the reflectivity of the unmodulated metasurface was measured at a 10º incidence angle with *p*-polarized light with all varactors uniformly biased (Supplementary Fig. 3). These measurements were carried out with broadband antennas as the source and receiver, and the VNA was used for a conventional S21 measurement with the source and receiver swept across the band of interest (5.5 - 8 GHz). We then characterized the spatio-temporal modulated metasurface. For these measurements the source (6.3 mW input power) is in continuous wave (CW) operation at 6.9 GHz with a 100 Hz resolution, and the detector was swept in frequency with corresponding 100 Hz resolution to measure the generated frequency harmonics. For beam steering measurements both the source and the detector were broadband horn antennas; however, for the focusing experiments a monopole antenna was used instead of one of the horn antennas. In the forward experiments the source (CW at $\omega_{in}$) was placed on-axis with the metasurface normal ($x$=0, $y$=0, $z$ =1.3) m, and the receiver was placed on a computer-controlled gimbal which scanned the reflected power in the $x$-$z$ plane. In the beam steering case this yielded radiation patterns of the harmonics scanned along a constant radius (Supplementary Fig. 1b); for focusing experiments a large portion of the $x$-$z$ plane was scanned yielding a 2D map of forward radiated power in the harmonics (Supplementary Fig. 1d). For the reverse nonreciprocity experiments the detector and source are swapped, and the new source operates in continuous wave operation at $\omega_{+1}$ (Supplementary Figs. 1c,e). The detector measures a frequency sweep as explained above, centered this time at $\omega_{in}$, capturing harmonics generated around this frequency. The receiver is again scanned in angle as in the forward experiment to measure reflected power over the area of interest.

**Theory.** The reflected field by the STMM is described by means of a generalized Bloch-Floquet approach based on a derivative expansion of the spatial phase distributions. It is given by Eq. (1), where $\widetilde{\mathcal{E}}_{\xi,n}[\boldsymbol{k}_{in}, \boldsymbol{k}_n, \omega_n; \boldsymbol{r}]$ is defined in Eq. (2) with $\Lambda_{\xi,n} = [R_{\xi,n-\text{sg}(n)}(\boldsymbol{k}_{in}, \omega_{in}; \boldsymbol{r}) \pm \delta_{|n|,1}]/[R_{\xi,n-\text{sg}(n)}(\boldsymbol{k}, \omega_{in}; \boldsymbol{r}) \pm \delta_{|n|,1}]$. Here, the '+' and '−' signs correspond to $\xi = s$ and $\xi = p$ polarizations, respectively. The reflection coefficients $R_{\xi,n}(\boldsymbol{k}, \omega_{in}; \boldsymbol{r})$ are obtained by using the local derivative expansion in Maxwell equations and solving numerically (or via continued fractions) an infinite set of coupled equations $\eta^+_{\xi,n-1} \mathcal{R}_{\xi,n-1} + A_{\xi,n} \mathcal{R}_{\xi,n} + \eta^-_{\xi,n+1} \mathcal{R}_{\xi,n+1} = 2 B_{\xi,n} \delta_{n,0}$. For the $p$-polarization, $\mathcal{R}_{p,n}(\boldsymbol{k}, \omega_{in}; \boldsymbol{r}) = R_{p,n}(\boldsymbol{k}, \omega_{in}; \boldsymbol{r}) - \delta_{n,0}$, $\eta^\pm_{p,n} =$

$\mp i\, (Z_s/2)(k_{zv,n}/k_{v,n})\, \Delta\sigma(\omega_n)\, e^{\pm i\,[\varphi(r)-r\cdot\nabla\varphi(r)]}$, $A_{p,n} = Z_s\, \sigma_{op}(\omega_n)\, (k_{zv,n}/k_{v,n}) +$ $i\, (k_{zv,n}\, k_{s,n}/k_{v,n}\, k_{zs,n})\, \cot(h\, k_{zs,n}) + Z_s/Z_v$, and $B_{p,n} = -Z_s/Z_v$. In these expressions, $Z_s$ is the spacer impedance, $h$ is the thickness of the spacer, $k_{v,n} = \omega_n/c$ and $k_{zv,n} = \sqrt{k_{v,n}^2 - k_n^2}$ are the magnitude of the wave-vector and its $z$-component in vacuum for the $n$-th harmonic, and similarly $k_{s,n} = k_{v,n}/Z_s$ and $k_{zs,n} = \sqrt{k_{s,n}^2 - k_n^2}$ are the corresponding quantities in the spacer. The reverse scattered field at frequency $\omega_{in}$ is given by a double integral over incoming and outgoing plane-wave components $\boldsymbol{E}_\xi^{\text{scatt,REV}}(\boldsymbol{r},z,t) = E^{in} \sum_{n=-\infty}^{\infty} \int \frac{d^2k'}{(2\pi)^2} \int \frac{d^2k}{(2\pi)^2} \hat{\boldsymbol{e}}_\xi^+(\boldsymbol{k}'_{-n})$ $\times\, \widetilde{\mathcal{E}}_{\xi,-n}[-\boldsymbol{k}_n, \boldsymbol{k}'_{-n}, \omega_{in}; \boldsymbol{r}]\, \widetilde{\mathcal{E}}_{\xi,n}[\boldsymbol{k}_{in}, \boldsymbol{k}_n, \omega_n; \boldsymbol{r}]\, e^{i\,[\boldsymbol{r}\cdot(\boldsymbol{k}'_{-n}-\boldsymbol{k}_n)+z\, k'_{zv,-n}-\omega_{in}t]}$. The scanning angles $\alpha_\pm^{\text{th}}$ at which the peaks in the reverse nonreciprocal focusing experiment are detected (Fig. 4c,d) are solutions to $\cos(\alpha_\pm^{\text{th}}) \tan(\alpha_\pm^{\text{th}} + g_\pm) - \sin(\alpha_\pm^{\text{th}}) = -2d/L_x$, where $g_\pm = \arccos[1 \pm (\omega_{in}/c)^{-1}(d/dx)\varphi_{n=+1}^{\text{focus}}(\pm L_x/2)]$. Here, $L_x$ is the size of the STMM along the $x$-direction, $d$ is the distance between the detector and the centre of the STMM.

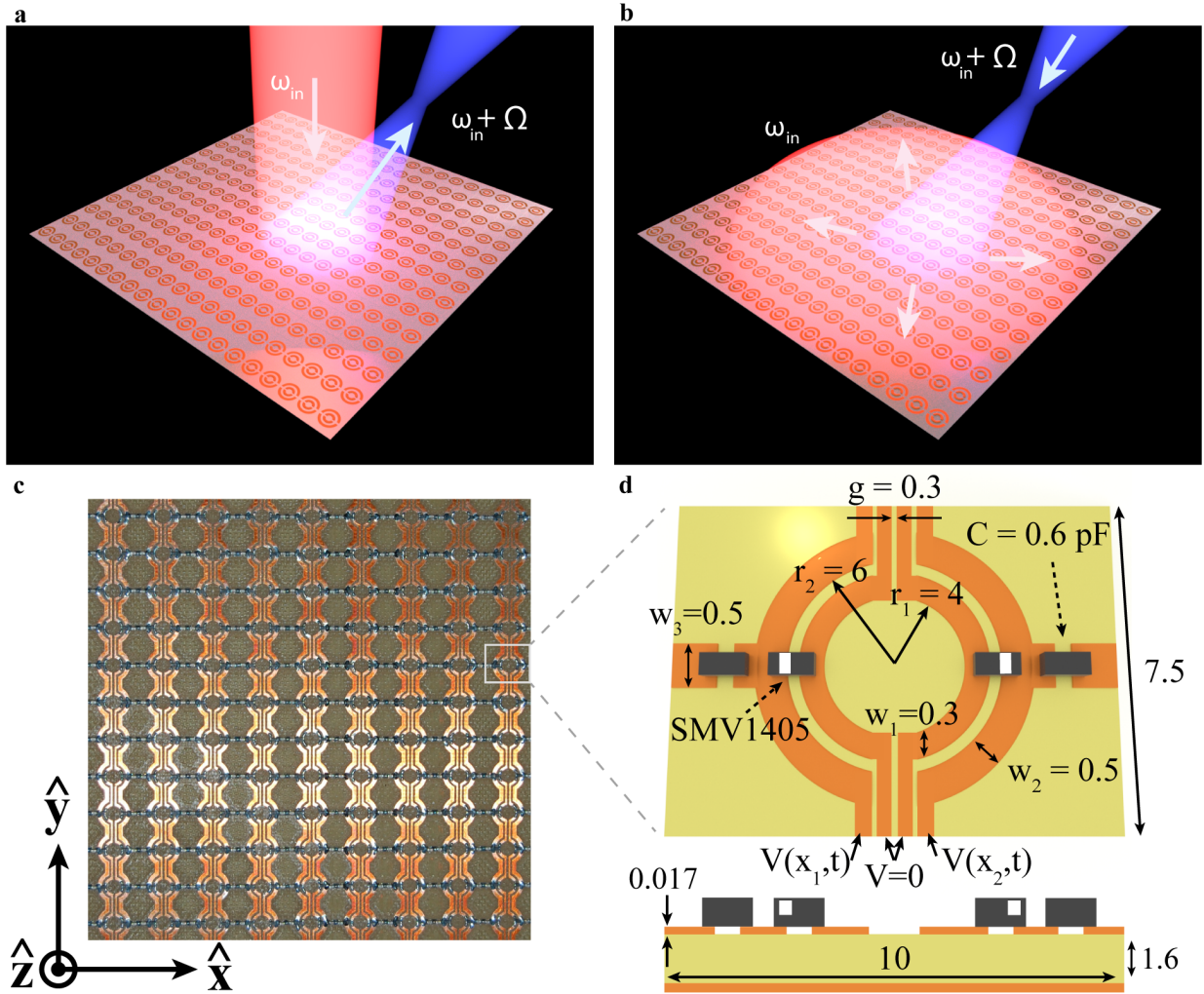

**Fig. 1 | Spatio-temporally modulated metasurfaces. a,** An incident beam impinging on an STMM is converted into a different harmonic that can be focused at any desired focal point. **b,** Breakdown of Lorentz reciprocity can be shown by probing the time-reversed process. **c,** Photograph of our STMM. **d,** Top view and cross-section of the unit cell. All geometrical parameters are in mm.

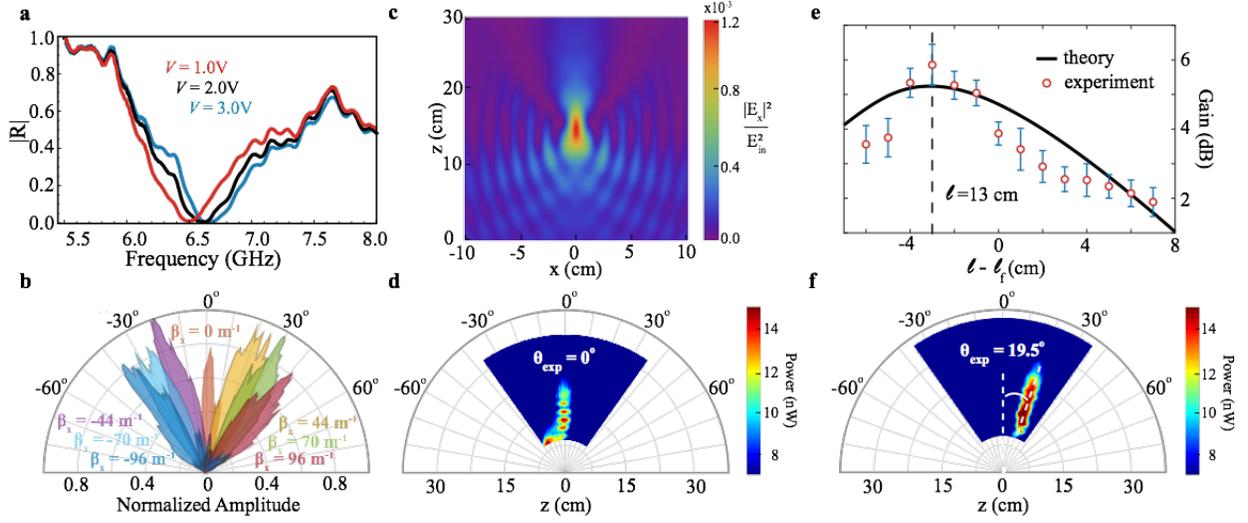

**Fig. 2 | Dynamical wave-front shaping. a,** Reflectance measurements for the unmodulated metasurface. **b,** Measured on-demand dynamical beam steering of the $n = +1$ harmonic. **c,** Calculated spatial distribution of the electric field for on-axis focusing for the $n = \bar{n} = +1$ frequency harmonic (infinite-sized STMM). **d,** Measured power for on-axis focusing. **e,** Calculated (finite-sized STMM, $L_x = 19$ cm) and measured gain as a function of $\ell$ for off-axis focusing. Error bars are determined by the standard deviation of gain over a narrow angular range within the focal region. **f,** Measured power for off-axis focusing. In b-f, $\boldsymbol{k}_{\text{in}} = 0$, $\omega_{\text{in}} = 2\pi \times 6.9$ GHz, and $\Omega = 2\pi \times 50$ kHz. Input power in the experiments is 6.3 mW. Focusing parameters are $(x_f, z_f) = (0,15)$ cm and $(6,15)$ cm for on- and off-axis, respectively.

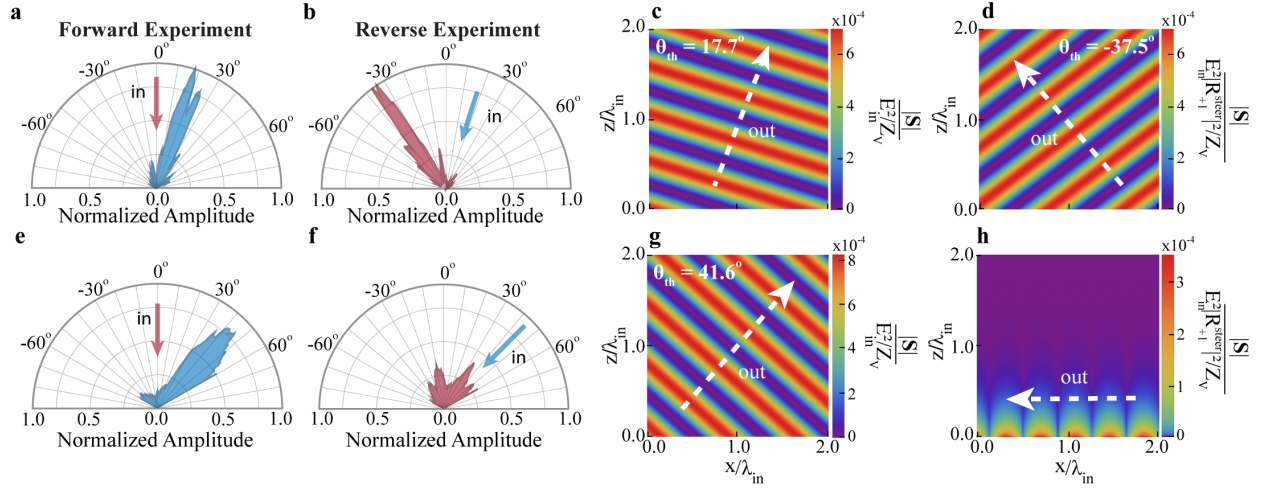

**Fig. 3 | Extreme nonreciprocity in beam steering. a,** An incident beam is steered to the $n = +1$ harmonic at an angle $\theta_{+1} \approx +18°$ with a phase gradient $\beta_x = 44$ m$^{-1}$. **b,** Reflection from $+18°$ of the up-converted beam undergoes a down-conversion to $\omega_{in}$ and is steered to $\approx -36°$. **c,** Calculated snapshot of the reflected Poynting vector distribution for the forward process in a. **d,** Same for the reverse process shown in b. **e,** Forward reflection to $\theta_{+1} \approx +43°$ for $\beta_x = 96$ m$^{-1}$. **f,** Reverse reflection to $\omega_{in}$ results in extreme breakdown of Lorentz reciprocity by launching surface waves. **g,** Computed reflected Poynting vector distribution for the forward process in e. **h,** Calculated profile of propagating surface waves corresponding to the reverse process in f.

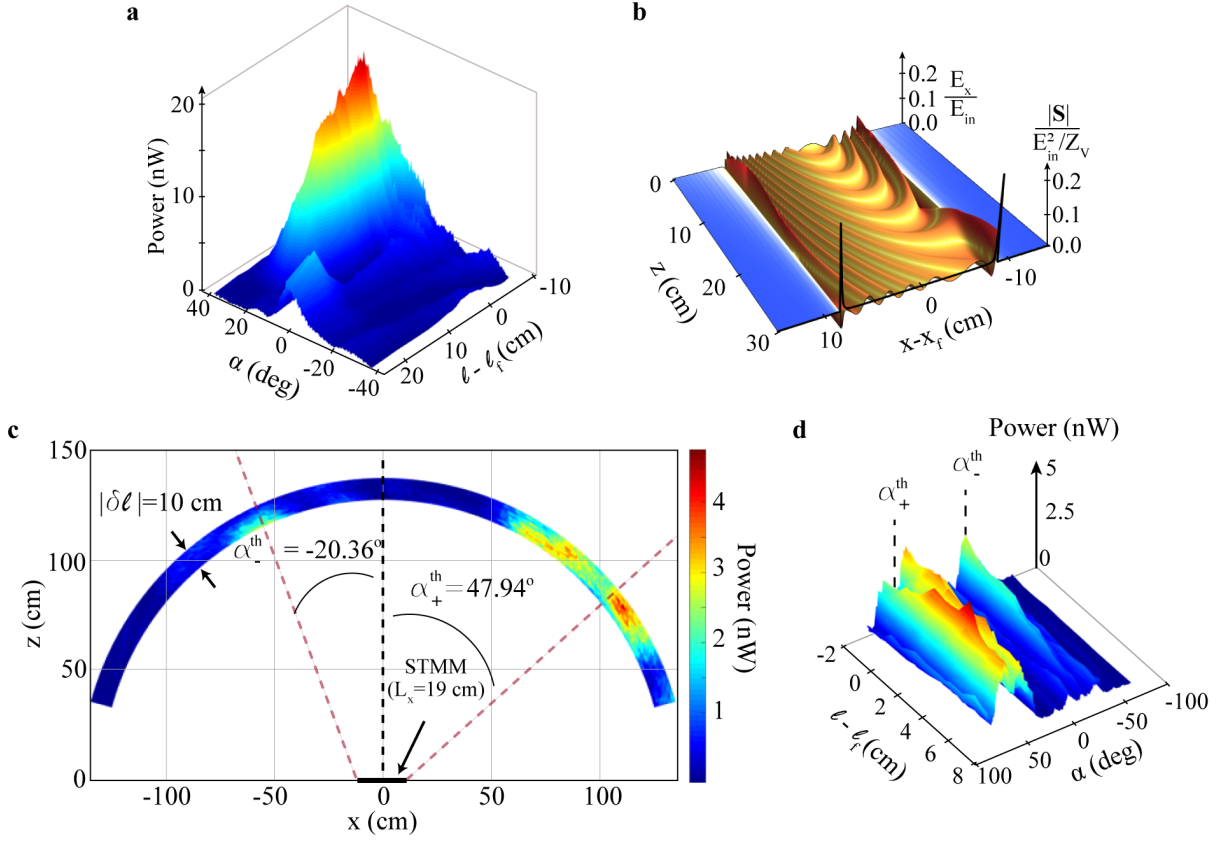

**Fig. 4 | Giant optical isolation in nonreciprocal focusing. a,** Measured reflected field power at $\omega_{+1}$ in the forward off-axis focusing experiment. **b,** Calculated snapshot of the reflected electric field distribution for the reverse scattering process $(-\boldsymbol{k}_{\text{in}} - \nabla\varphi_{\bar{n}=+1}^{\text{focus}}, \omega_{\text{in}} + \Omega) \rightarrow (-\boldsymbol{k}_{\text{in}} - 2\nabla\varphi_{\bar{n}=+1}^{\text{focus}}, \omega_{\text{in}})$ for the infinite-sized STMM. The black line represents the time-averaged Poynting vector for the same scattering process. **c,** Measured power at $\omega_{\text{in}}$ in the far-field for the reverse process is mainly confined within the "wedge"-shaped region. Data at given radii of the arc sector correspond to different locations $\ell$ of the monopole source. **d,** Same as c but as a function of the scanning angle and $\ell$. Parameters are $x_f = 6$ cm and $z_f = 15$ cm ($\ell_f = 16$ cm).

24